\documentclass{article}

\usepackage{PRIMEarxiv}

\usepackage[utf8]{inputenc} 
\usepackage[T1]{fontenc}    
\usepackage{hyperref}       
\usepackage{url}            
\usepackage{booktabs}       
\usepackage{amsfonts}       
\usepackage{nicefrac}       
\usepackage{microtype}      
\usepackage{lipsum}
\usepackage{fancyhdr}       
\usepackage{graphicx}       
\graphicspath{{media/}}     

\title{Preserving the Ephemeral: \textit{Instagram} Story Archiving with the \textit{Tidal Tales Plugin}
}

\author{
  Michael Achmann-Denkler, Christian Wolff \\
  Media Informatics Group \\
University of Regensburg \\
Regensburg, Germany \\ 
  \texttt{\{michael.achmann, christian.wolff\}@ur.de} \\
}

\begin{document}
\maketitle

\begin{abstract}
We introduce the \textit{Tidal Tales Plugin}, a Firefox extension for efficiently collecting and archiving of \textit{Instagram} stories, addressing the challenges of ephemeral data in social media research. It enables an automated collection of story metadata and media files without risking account bans. It contributes to Web Science by facilitating expansive, long-term studies with enhanced data access and integrity.
\end{abstract}


\section{Background}
\textit{Instagram} stories are an ephemeral social media format. This special format was introduced to the platform in 2016 and is attributed as ``the biggest driver for \textit{Instagram}'s overall success'' \cite{Leaver2020-cj}. With a growing body of literature about \textit{Instagram} content \cite{Rejeb2022-xr}, one would expect a plethora of studies regarding stories. Yet, most projects concentrate on permanent posts, limiting their data collection to this type of content. Some studies acknowledge this limitation without detailing the reasons for the shortcomings. Through our experience collecting stories, we identified two major obstacles: 1) the ephemeral character calls for prospective data collection and 2) a lack of suitable scraping services or tools. Posts, in contrast, can be collected retrospectively, offering access to large corpora within seconds. The \textit{Meta Content Library}\footnote{Until August 2024: CrowdTangle.}, a tool open to researchers, facilitates access to public Instagram posts, offers search and filter functions, and allows the export of posts and their metadata. Contrasting the data accessibility of the two formats suggests that researchers might follow the path of least resistance for data collection, possibly introducing a bias \cite{Trezza2023-vk}.


To overcome the obstacles to collecting Instagram stories, we propose the \textit{Tidal Tales Plugin}, a Firefox extension to collect ephemeral stories and their metadata. The project is a fork of \textit{Zeeschuimer} \cite{stijn_peeters_2023_8399900} and facilitates the prospective data collection of both, private as well as public \textit{Instagram} accounts. All software is shared under open-source licenses, enabling researchers to set up their projects to meet privacy and data safety requirements. We aim at a low level of technical proficiency, after a one-time setup data can be collected and exported without technical knowledge. 

\subsection{Data Collection in the Literature}
A small collection of studies from different backgrounds addresses stories, each using a unique approach to data collection. The most straightforward approach to collecting a story corpus is taking screenshots, as exercised for a master's thesis \cite{Amancio2017-ol}. Some studies do not go into detail on their data collection, merely stating that ``[s]tories were downloaded manually into an Excel file'' \cite{Towner2024-gr} or -- even more generic -- that they were collected \cite{Li2021-mm}. Two studies relied on tools: \textit{4K Stogram} was used to collect stories during the 2020 U.S. presidential campaign \cite{Towner2022-oe} and \textit{StorySaver} for the collection of private stories \cite{Bainotti2020-rn}. One study reported the use of \textit{Selenium} for scraping \cite{Achmann2023-rs}, and another study the use of a \textit{Chrome} plugin without any further specifications \cite{Vazquez-Herrero2019-ue}. Some publications report the time of data collection, usually once within 24 hours at a set time. The amount of collected story items ranges from 304 \cite{Towner2022-oe} to 2208 \cite{Achmann2023-rs}, the data collection phase ranges from 7 days \cite{Bainotti2020-rn} to three weeks at a time \cite{Amancio2017-ol}. 










\section{\textit{Tidal Tales Plugin}}
The \textit{Tidal Tales Plugin} is a \textit{Firefox} extension to collect stories while browsing.\footnote{See documentation: \href{https://tidaltal.es}{https://tidaltal.es}. Source code available at: \href{https://github.com/michaelachmann/tidal-tales-plugin}{https://github.com/michaelachmann/tidal-tales-plugin}} In contrast to previous work, we suggest collecting stories twice daily, mirroring the natural rhythm of tides, to eliminate the risk of overlaps and to add a margin for error (i.e., time to fix problems). The plugin justifies doubling the work: Once researchers have created an \textit{Instagram} account to follow their target accounts, they only need to open the webpage and access the first story.\footnote{This process may be automated using \textit{Selenium}.} The plugin enables long-term data collection without restrictions on the number of target accounts.



\begin{figure}
    \centering
    \includegraphics[width=.75\linewidth]{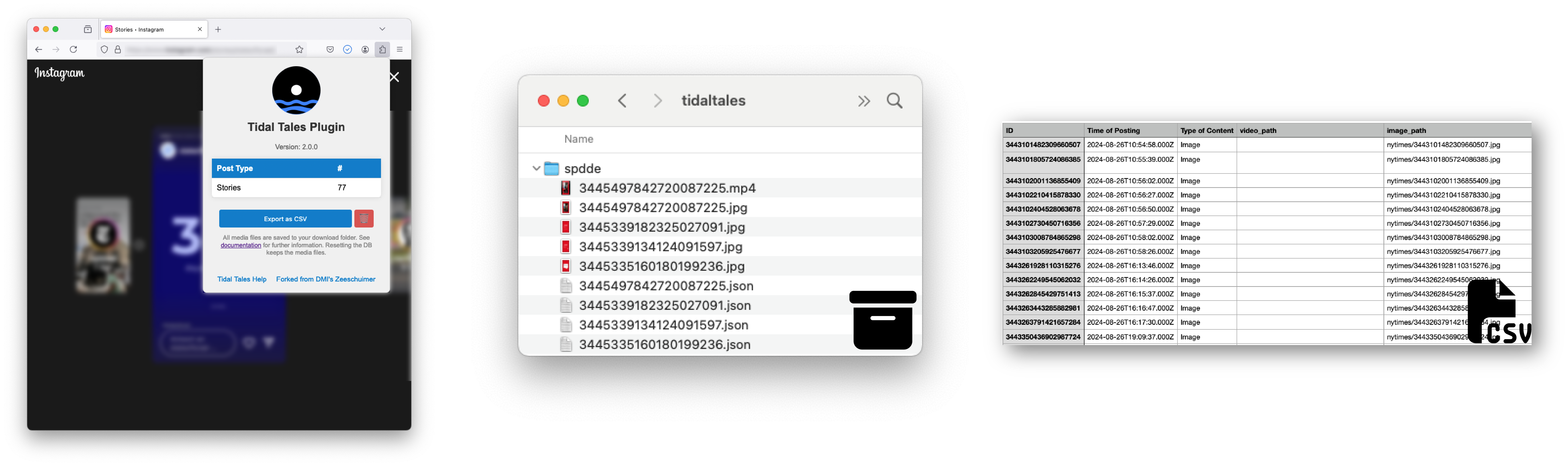}
    \caption{A screenshot of the tool: The plugin runs inside Firefox. When watching stories, media files and metadata are downloaded in the background. All files are saved to the local file system. Additionally, the metadata can be exported as a CSV file.}
    \label{fig:enter-label}
\end{figure}

The plugin is a modified fork of Stijn Peeters's \textit{Zeeschuimer} \cite{stijn_peeters_2023_8399900}. The original software was developed for use with \textit{4CAT} \cite{Peeters2022-lp}. The plugin uses the \textit{Firefox} API to intercept background requests between the \textit{Instagram} application in the browser and its server. When opening and browsing stories, the plugin detects the corresponding requests and logs the metadata in the database. Next, videos and images are downloaded through Firefox. This approach has two main advantages:
\begin{enumerate}
    \item It is minimally invasive to stay within the legal constraints for data mining in a scientific context: The plugin captures data during normal usage. Researchers use the website as intended -- except that files and information shared within normal usage are additionally stored on the local file system and a local database inside Firefox.\footnote{German §60d \textit{Urhberrechtsgesetz} \cite{urhg_60d} contains exceptions for non-profit researchers. The European Directive 2019/790 \cite{directive_2019_790_art3} similarly provides provisions for text and data mining activities. Researchers must consult local laws and institutional guidelines before using Tidal Tales.} 
    \item By storing files locally, researchers have full control over the collected data and can implement their research and institution's ethics and data privacy guidelines. 
\end{enumerate}

    


\section{Conclusion}

We introduced \textit{Tidal Tales Plugin}, a tool to streamline the collection and archiving of Instagram stories. Our solution effectively solves several challenges: by enabling the collection of stories directly within the original web application during browsing and viewing, it lowers the risk of account bans -- a frequent complication with tools like \textit{Instaloader}. Unlike commercial services such as \textit{StorySaver} and \textit{4K Stogram}, \textit{Tidal Tales} captures detailed story metadata, this feature allows the direct analysis of platform affordances. The automation of our approach permits long-term studies with larger sample sizes than previously feasible. 

Furthermore, while stories would dictate prospective data collection, Instagram's highlight feature presents an opportunity for researchers to access selected expired stories. This feature allows profile authors to curate and showcase collections of ephemeral posts in organized highlights. Our plugin can capture these highlighted stories during browsing, extending the scope of data collection beyond the typical 24-hour lifespan. 

The practice of scraping is considered a ``necessary evil'' by Venturini and Rogers, who also recommend a conscientious approach to data collection \cite{Venturini2019-ve}. In our context, it is important to understand that users post stories with the expectation of their transient nature. Therefore, researchers should obtain consent and follow guidelines set forth by ethics committees when collecting stories, particularly private ones. 

\subsection{Future Work}
We are considering adding anonymization and pseudonymization functionalities to enhance data protection in a future plugin version.





\bibliographystyle{unsrt}  
\bibliography{references}

\end{document}